%% file: main.tex
\documentclass[preprint,journal]{vgtc}       

\ifpdf
  \pdfoutput=1\relax                   
  \usepackage{graphicx}                
  \DeclareGraphicsExtensions{.pdf,.png,.jpg,.jpeg} 
\else
  \usepackage{graphicx}                
  \DeclareGraphicsExtensions{.eps}     
\fi%

\graphicspath{{figures/}{pictures/}{images/}{./}} 

\usepackage{microtype}                 
\PassOptionsToPackage{warn}{textcomp}  
\usepackage{textcomp}                  
\usepackage{mathptmx}                  
\usepackage{times}                     
\usepackage{cite}                      
\usepackage{tabu}                      
\usepackage{booktabs}                  

\usepackage[normalem]{ulem}

\newif\ifnotes
\notesfalse 

\onlineid{1051}

\vgtccategory{Research}
\vgtcpapertype{system}

\title{Lyra 2: Designing Interactive Visualizations by Demonstration}

\author{Jonathan Zong, Dhiraj Barnwal, Rupayan Neogy, and Arvind Satyanarayan}
\authorfooter{
\item
 Jonathan Zong, Rupayan Neogy, and Arvind Satyanarayan are with the Massachusetts Institute of Technology. E-mails: jzong@mit.edu, rneogy@mit.edu, arvindsatya@mit.edu.
\item
 Dhiraj Barnwal is with Indian Institute of Technology Kharagpur. E-mail: sumankumaribarnwal@gmail.com.
}

\shortauthortitle{Zong \MakeLowercase{\textit{et al.}}: Lyra 2: Designing Interactive Visualizations by Demonstration}

\input{sections/abstract.tex}

\CCScatlist{ 
 \CCScat{H.5.2}{INFORMATION INTERFACES AND PRESENTATION}%
{User Interfaces}{Interaction styles};
}

\teaser{
  \centering
  \includegraphics[width=0.95\linewidth]{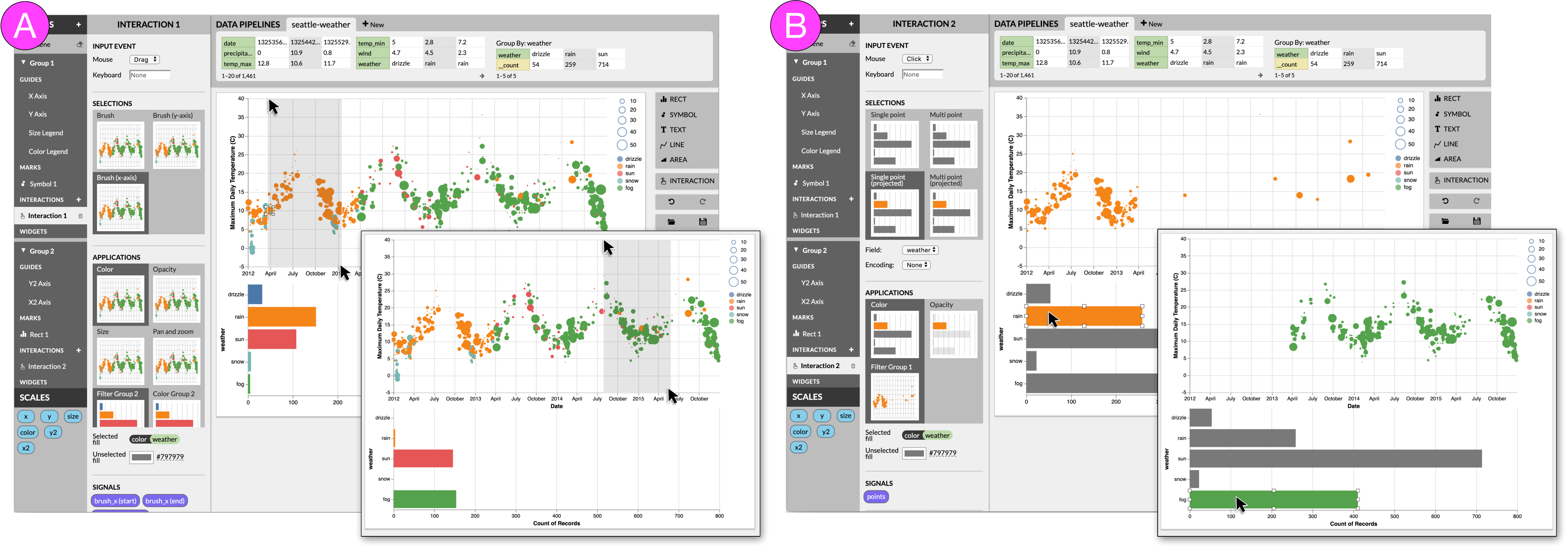}
  \vspace{-3mm}
  \caption{An example interactive visualization designed in Lyra 2, a visualization design environment. Users can (a) brush in the scatterplot to re-aggregate the histogram, and (b) click histogram bars to filter for corresponding points in the scatterplot. This visualization was designed by demonstration\,---\,users did not have to write any textual code.}
  \label{fig:teaser}
}

\vgtcinsertpkg

\begin{document}


\firstsection{Introduction}

\maketitle

\definecolor{purplecolor}{RGB}{128, 0, 128}
\definecolor{grey}{RGB}{128, 128, 128}

\DeclareRobustCommand{\hsout}[1]{\texorpdfstring{\renewcommand{\cite}{\ccite}\sout{#1}}{#1}}
\DeclareRobustCommand{\del}[1]{{\ifnotes{\leavevmode\color{grey}{\protect\hsout{#1}}}\fi}}
\newcommand{\add}[1]{\ifnotes{\color{purplecolor}{#1}}\else{#1}\fi}
\newcommand{\replace}[2]{\ifnotes{\del{#1}\add{#2}}\else{#2}\fi}


\input{sections/intro}
\input{sections/related_work}
\input{sections/background}
\input{sections/system_design}
\input{sections/eval_examples}

\input{sections/eval_study}
\input{sections/eval_cognitive}
\input{sections/conclusion}
\acknowledgments{
We thank our study participants and anonymous reviewers for their invaluable feedback, and the MIT Visualization Group for their camaraderie.
We are also grateful to the people who helped rebuild the Lyra infrastructure in 2016 including Jeffrey Heer, who supported the effort via his Moore Foundation award, and K. Adam White and Sue Lockwood who led the work. 
This project was supported by the Paul and Daisy Soros Fellowship and NSF Award \#1942659.}

\bibliographystyle{abbrv-doi}

\bibliography{main}
\end{document}

%% file: sections/abstract.tex
\abstract{
Recent graphical interfaces offer direct manipulation mechanisms for authoring visualizations, but are largely restricted to static output. To author interactive visualizations, users must instead turn to textual specification, but such approaches impose a higher technical burden. To bridge this gap, we introduce Lyra 2, a system that extends a prior visualization design environment with novel methods for authoring interaction techniques by demonstration. Users perform an interaction (e.g., button clicks, drags, or key presses) directly on the visualization they are editing. The system interprets this performance using a set of heuristics and enumerates suggestions of possible interaction designs. These heuristics account for the properties of the interaction (e.g., target and event type) as well as the visualization (e.g., mark and scale types, and multiple views). Interaction design suggestions are displayed as thumbnails; users can preview and test these suggestions, iteratively refine them through additional demonstrations, and finally apply and customize them via property inspectors. We evaluate our approach through a gallery of diverse examples, and evaluate its usability through a first-use study and via an analysis of its cognitive dimensions. We find that, in Lyra 2, interaction design by demonstration enables users to rapidly express a wide range of interactive visualizations.
} 

\keywords{Direct manipulation, interactive visualization, interaction design by demonstration}

%% file: sections/intro.tex
Interactive visualization is increasingly embraced as a medium for recording, analyzing, and communicating data. 
To meet this demand, a recent thread of research has explored methods for minimizing the technical expertise required to author visualizations.
Systems like Lyra~\cite{satyanarayan2014lyra}, Data Illustrator~\cite{liu2018data}, and Charticulator~\cite{ren2018charticulator} provide graphical interfaces for creating visualizations with drag-and-drop and direct manipulation interactions rather than programming.
Though interactivity is recognized as crucial to effective visualization~\cite{pike:interactionscience, liu:mentalmodels}, few graphical interfaces offer support for interaction design\,---\,the aforementioned systems only produce static output, and other alternatives including Tableau (née Polaris~\cite{stolte2002polaris}) and VisDock~\cite{choi2015visdock} either hard-code specific interaction techniques, or offer only a limited typology to choose from.

To author \emph{custom} interactive visualizations, users must instead turn to textual specification languages, such as D3~\cite{bostock2011d3}, Vega~\cite{satyanarayan2014declarative, satyanarayan2015reactive}, and Vega-Lite~\cite{satyanarayan2016vegalite}. 
While highly expressive, these tools have several usability drawbacks compared to graphical interfaces.
For instance, with D3, authors must write low-level event callbacks which expose execution details like mutable state and concurrency~\cite{cooper:embedding, edwards:coherent}.
These details are often unrelated to visualization design, and managing them hinders authors from quickly iterating on designs. 
Declarative languages such as Vega and Vega-Lite have made progress by introducing higher-level abstractions to mask these execution concerns.
However, these abstractions are expressed through textual specifications which present an unnecessarily large \emph{gulf of execution}~\cite{hutchins1985direct} by providing a poor \emph{closeness of mapping}~\cite{blackwell:cogdim} to the ultimate interactive visual output.
As a result, users are forced to learn and juggle two very different paradigms.

To bridge this gap, we introduce \add{Lyra 2, a system that extends a prior visualization design environment~\cite{satyanarayan2014lyra} with} methods for designing interactive visualizations by demonstration.
Consider the example task of creating a rectangular brush for selecting and highlighting points on a scatterplot. 
To specify this interaction, users demonstrate it by dragging their mouse cursor directly over the visualization they are currently editing.
\replace{The system}{Lyra 2} interprets this performance using heuristics, and suggests possible interaction techniques to apply.
In our example, the system detects the drag events in a space marked by quantitative x- and y-axes and suggests a set of interval-based interactions~\cite{satyanarayan2016vegalite}.
Suggestions consist of a selection (e.g., 1D or 2D brushes) and an application (e.g., conditional color or opacity encodings, or filter transforms). 
Suggestions are displayed as thumbnail previews, which facilitate rapid comparison by illustrating what the visualization would look like after applying the interaction.
Users can perform additional demonstrations to refine the suggestions, or click to accept a suggested interaction.

\replace{We instantiate this approach in Lyra 2, building on an existing visualization design system~\cite{satyanarayan2014lyra}, with demonstrations and suggestions generating}{In Lyra 2, demonstrations and suggestions generate} statements in \del{the }Vega~\cite{satyanarayan2015reactive} or Vega-Lite~\cite{satyanarayan2016vegalite}\del{ visualization grammars}.
Critically, our approach smooths the gradient of these two levels of abstraction. 
For instance, say we wished to label the corners of a rectangular brush with their data coordinates. 
Vega-Lite does not provide any facilities to do so; a user could choose to edit the compiled Vega specification to add the appropriate signals, but would experience a sharp complexity cliff and have to reason about two saliently different paradigms (selections and reactive programming, respectively). 
Demonstrations and suggestions, however, provide a consistent interface mechanism to seamlessly move between these two levels of abstraction.
Once a user demonstrates a brush interaction, they can use visual property inspectors to drill into the components of the interaction (i.e., the brush start and end extents); these extents can then be dropped over text mark properties to achieve the desired behavior. 

To evaluate our approach, we follow current best practices~\cite{ren2018reflecting} by using three distinct evaluation methods. 
To assess its expressive extent, we use demonstrations in Lyra 2 to recreate a diverse gallery of examples that highlight substantial coverage of an existing taxonomy of interaction techniques for data visualizations~\cite{yi:understanding}.
To determine its usability, we conducted a first-use study with participants spanning a broad range of prior experience creating interactive visualizations. 
All study participants were able to recreate a range of interactive visualizations and described the use of demonstrations as ``natural''.
Finally, we analyze the cognitive dimensions~\cite{blackwell:cogdim} of our approach to further assess usability, and find that demonstrations and suggestions offer a means to \emph{progressively evaluate} desired interactive outcomes with a much \emph{closer mapping} between the specification and output medium.

%% file: sections/related_work.tex
\begin{figure*}[t]
  \centering
  \includegraphics[width=0.95\linewidth]{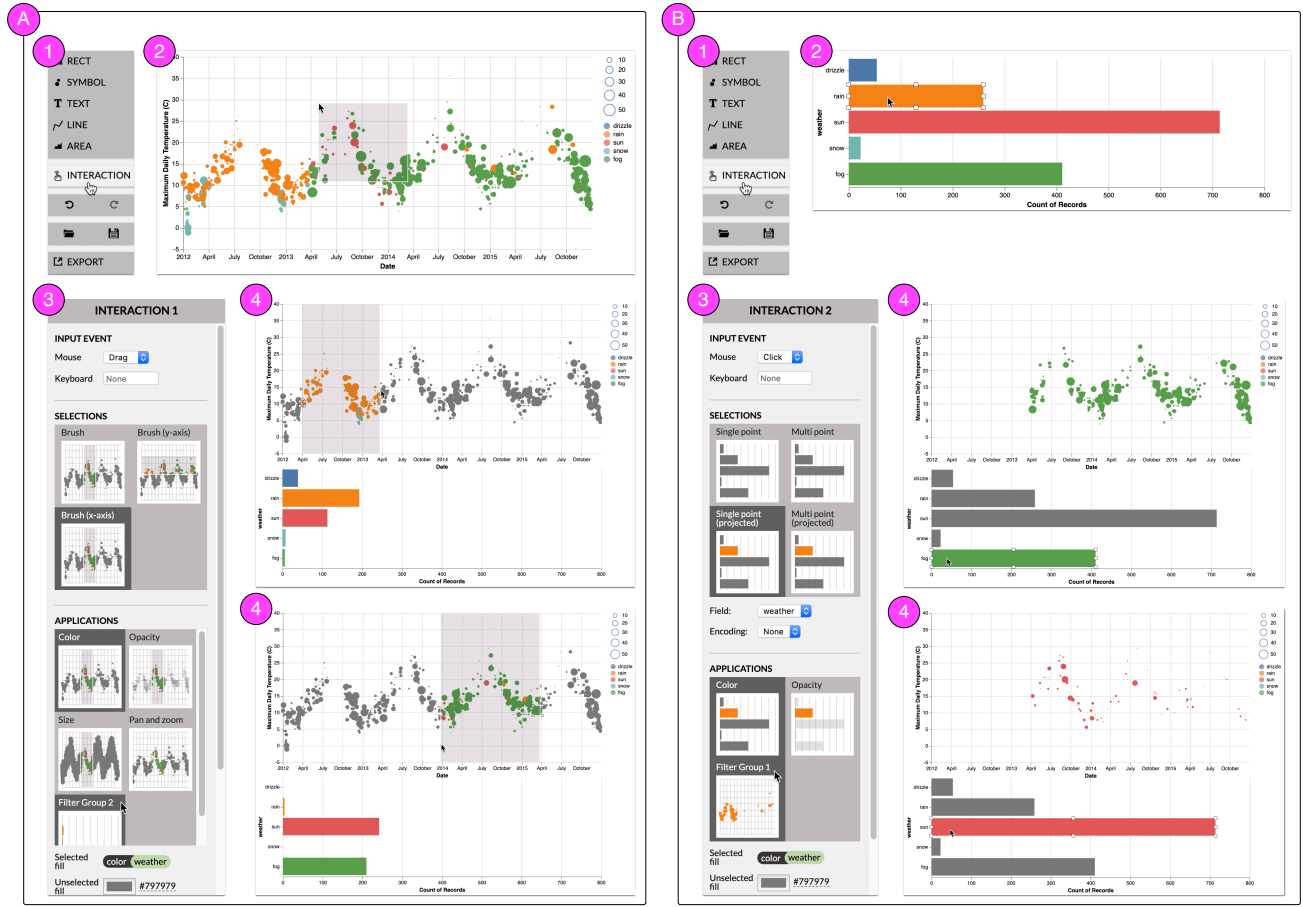}
  \vspace{-3mm}
  \caption{The process of creating the interactive visualization in Fig. \ref{fig:teaser}. (a) Demonstrating a horizontal brush that colors selected points in the scatterplot and re-aggregates the histogram. (b) Demonstrating a click interaction that colors the selected bar and filters the scatterplot via the \texttt{weather} field.}
  \label{fig:walkthrough}
  \vspace{-5mm}
\end{figure*}

\section{Related Work}

Our contribution builds on prior work on models of interactive visualization design and graphical interfaces for authoring visualizations, \replace{including recent work on visualization by demonstration}{and is} inspired by \add{the literature on} programming by demonstration (PBD).

\subsection{Textual Specification of Interactive Visualizations}

A variety of textual specification languages and visualization toolkits have explored methods for specifying custom interactive behaviors.
Protovis~\cite{2009-protovis}, D3~\cite{bostock2011d3}, \add{and VisDock~\cite{choi2015visdock}}, for example, offer palettes of standard techniques but force users to write low-level imperative event handling for custom behaviors.
Improvise~\cite{weaver2004improvise} and Stencil~\cite{cottam2008stencil} offer more fine-grained primitives, inspired by data flow semantics, that dynamically update and propagate values to downstream dependents\,---\,a conceptual model that allows for expressive interaction design.
More recently, Vega~\cite{satyanarayan2015reactive, satyanarayan2014declarative} and Vega-Lite~\cite{satyanarayan2016vegalite}, which we describe in greater detail in Section \ref{sec:background}, explore grammar-based approaches for specifying interaction techniques.
While these tools span the gamut of expressivity, they share common usability concerns.
Namely, they each present a non-trivial \emph{gulf of execution}~\cite{hutchins1985direct} because they force users to express interactive and visual outputs in terms of text\,---\,a mismatch between the input and output representations. 
As a result, although these tools have high expressive ceilings, they present a non-trivial threshold~\cite{myers2000past} and thus are typically favored by users with prior visualization expertise.

\subsection{Non-Textual Specification of Visualizations}

To make visualization design more accessible to users with less technical expertise, researchers have explored graphical interfaces for visualization design. 
Systems like Lyra~\cite{satyanarayan2014lyra}, Data Illustrator~\cite{liu2018data}, and Charticulator~\cite{ren2018charticulator} (three recent examples in a rich design space~\cite{kim2016data, mauri2017rawgraphs, mendez2016ivolver, mei2018design}) allow users to author visualizations through direct manipulation interactions inspired by vector graphics editors.
While these systems differently trade expressivity for learnability~\cite{2020-critical-reflections}, none yet explore specifying custom interaction techniques.
Graphical interfaces that do support interactivity typically either hard code specific interactions (e.g., \del{Tableau~\cite{stolte2002polaris} or }iVisDesigner~\cite{ren2014ivisdesigner} which only supports brushing \& linking) or allow users to instantiate behaviors from a predefined palette of techniques.
For example, both Microsoft PowerBI and Tableau allow users to create dynamic query widgets (called \emph{parameters} in both tools) or specify interactive filters and highlights (called \emph{actions} in Tableau). 
However, users are restricted to a handful of customizations (e.g., actions can only run on hover, selection, or via the context menu rather than the rich space of mouse and keyboard events), and interactive mechanisms can only be applied in a limited fashion (e.g., interactive filters can show or hide data but cannot drive downstream calculations).
Thus, enabling \add{more expressive,} custom interaction design for data visualization without textual programming remains an open problem.

\subsection{Programming by Demonstration (PBD)}

Research in programming by demonstration (PBD) and programming by example (PBE) has investigated how to interpret users' direct manipulation inputs to generate programs as output. For example, users can demonstrate string manipulation operations by providing input-output pairs~\cite{lau2003programming, gulwani2011automating}, can automate scraping by recording and replaying interactions with web pages~\cite{chasins2015browser, barman2016ringer}. Similarly, Data Wrangler~\cite{kandel2011wrangler} generates specifications of data transformations based on a user's direct manipulation of tabular data, and provides visual previews of possible outputs. These systems narrow the \textit{gulf of execution} by eliciting inputs using the same representation as the desired output.

Through systems like Gold~\cite{myers1994creating} and a recent thread by Saket et al.~\cite{saket2016visualization, saket2019investigating, saket2019liger}, researchers have shown that PBD is also a viable approach for designing visualizations.
With PBD, rather than explicitly binding data fields to encoding channels, users implicitly specify these mappings by performing demonstrations\,---\,for instance, when the user drags two points together, the system infers that they intend to create a scatterplot and suggests several x-y axis pairs.
Our approach shares similarities with this line of work: users perform demonstrations directly on the visualization, \add{which are interpreted by a series of rules to enumerate candidate choices} rendered as visual previews.\del{, and users can refine suggestions via additional demonstrations.}
\replace{But the two methods diverge in the outcome of the demonstration: with Saket et al.'s systems, users perform a demonstration to express how a static visualization should be constructed; with our approach, demonstrations express interaction techniques.}
{%
But salient differences arise due to the outcome of the demonstration. 
For instance, with Saket et al.'s systems, demonstrations produce static visualizations for visual data exploration.
As a result, system rules interpret demonstrations by mapping them to data analysis goals, and each suggestion has a computed relevance score~\cite{saket2019demonstrational}. 
In contrast, demonstrations in Lyra 2 specify custom interaction designs. 
Our heuristics use demonstrations to generate all valid statements in the underlying Vega or Vega-Lite visualization grammars, and then set defaults.%
}

Finally, our work draws inspiration from systems like Monet~\cite{li2005informal} and Peridot~\cite{myers1987creating}, which investigate how demonstrations can be used to construct interactive user interfaces.
In our domain, the Vega and Vega-Lite visualization grammars influence our system design (e.g., demonstrations can be processed with a simpler set of heuristics instead of Monet's neural-network-based algorithms).
But, they also provide an opportunity to more carefully analyze the usability tradeoffs of textual versus demonstrational specification of interaction techniques. 
For instance, Peridot provides an ``active value'' primitive, with facilities to individually remove associated interactions; these concepts map to signals and the affordances of Lyra 2's property inspectors.
However, with Peridot, users can never edit an equivalent textual specification of the interaction techniques (with Lyra 2, users can export the underlying Vega specification).
As a result, with Lyra 2, we are able to identify that property inspectors help reduce hidden dependencies that may exist in a textual specification, but yield a more diffuse user experience.





%% file: sections/background.tex
\section{Background}
\label{sec:background}

Our implementation of interaction design by demonstration in Lyra 2 builds on both its existing support of static visualization design~\cite{satyanarayan2014lyra}, and on interaction design concepts in Vega~\cite{satyanarayan2014declarative, satyanarayan2015reactive} and Vega-Lite~\cite{satyanarayan2016vegalite}.
In this section, we aim to provide the reader with sufficient background to understand the remainder of the paper. 

\subsection{Static Visualization Design in Lyra}

Lyra offers the following abstractions for authoring static data visualizations, inspired by a similar set of abstractions found in Vega. Data \emph{pipelines} allow users to import tabular datasets, inspect them via a data table view, and apply chains of\del{ common} statistical data transformations (e.g., filtering and grouping). \emph{Scales} map data values to visual properties such as position, shape, and color. Lyra supports both discrete and quantitative scales, and automatically instantiates an appropriate scale when a direct manipulation data binding operation occurs. \del{Users can tweak scale properties (or manually create a scale) using visual property inspectors. }\emph{Guides} are reference marks that visualize scales: axes visualize scales over a spatial domain and legends visualize scales for color, shape, or size encodings. Like scales, Lyra automatically constructs\del{ necessary} guides when a data bind occurs\del{, and users can manually construct them or adjust their properties via property inspectors}. \emph{Marks} are shapes (e.g. rectangles, lines, symbols, text labels) with named visual properties (e.g., x, y, and fill). Property values can be set to constant\replace{ values}{s}, or bound to data. When a mark definition is bound to a dataset, Lyra instantiates one mark instance per datum.

To directly manipulate \replace{the}{these} abstractions\del{ described above}, Lyra provides the following user interface components. Akin to vector graphics packages, marks appear on the \emph{visualization canvas} with \emph{handles}, which can be used to interactively move, rotate, and resize all instances of the selected mark. When dragging a field from a pipeline's data table\del{ view}, shaded regions called \emph{dropzones} overlay the canvas. Each dropzone represents a mark property, like color or x position. When the user drops a field onto these targets\del{ overlays}, Lyra binds that field to the mark property. The canvas always reflects the current state of the output visualization. \textit{Property inspectors} list features of \replace{marks, scales, guides, and data transformations}{the visualization's components}, and provide an interface for fine-grained editing\del{ as an alternative to the canvas}. \replace{Property values}{Properties} may \replace{be edited directly or}{also be} set \replace{via drag-and-drop of}{by dropping} data fields, and \replace{these}{any} changes are \replace{reflected}{shown} immediately on the canvas.

\subsection{Interaction Design in Vega-Lite and Vega}

To support interaction design by demonstration, Lyra builds on the following two abstractions provided by Vega-Lite and Vega respectively. In Vega-Lite, interaction techniques are expressed as \textit{selections}, which are sets of data records that a user has interacted with. Three types of selections are supported\,---\,single, multi, or interval\,---\,which determine the logic for which records are included within the set, and what input event triggers this inclusion (e.g., mouse clicks). Selections can be \emph{projected} to vary the inclusion criteria. For instance, a \del{default }single selection only includes the \del{specific }point a user clicked\del{ on}; projecting the selection over a field will include the clicked point and all points that share its value for \replace{the given}{that} field. Selections can \replace{be used to define}{drive} conditional encodings, which apply different values depending on whether \del{or not }a data record is selected. Selections can also \del{be used to }filter input data and\del{ to} determine scale domains. Vega-Lite selections compile into \replace{expressions involving Vega signals}{Vega signal expressions}. In Vega, \emph{signals} are dynamic variables which update in response to input event streams\del{ or changes to HTML widgets (e.g., range sliders)}. They can be composed into expressions, which are formulas that\del{ are} automatically recalculate\del{d} whenever signal values change. Signals can be used throughout a Vega specification, including as part of a mark property, data transform parameter, or scale domain.


%% file: sections/system_design.tex
\begin{figure*}[t]
  \centering
  \includegraphics[width=0.95\textwidth]{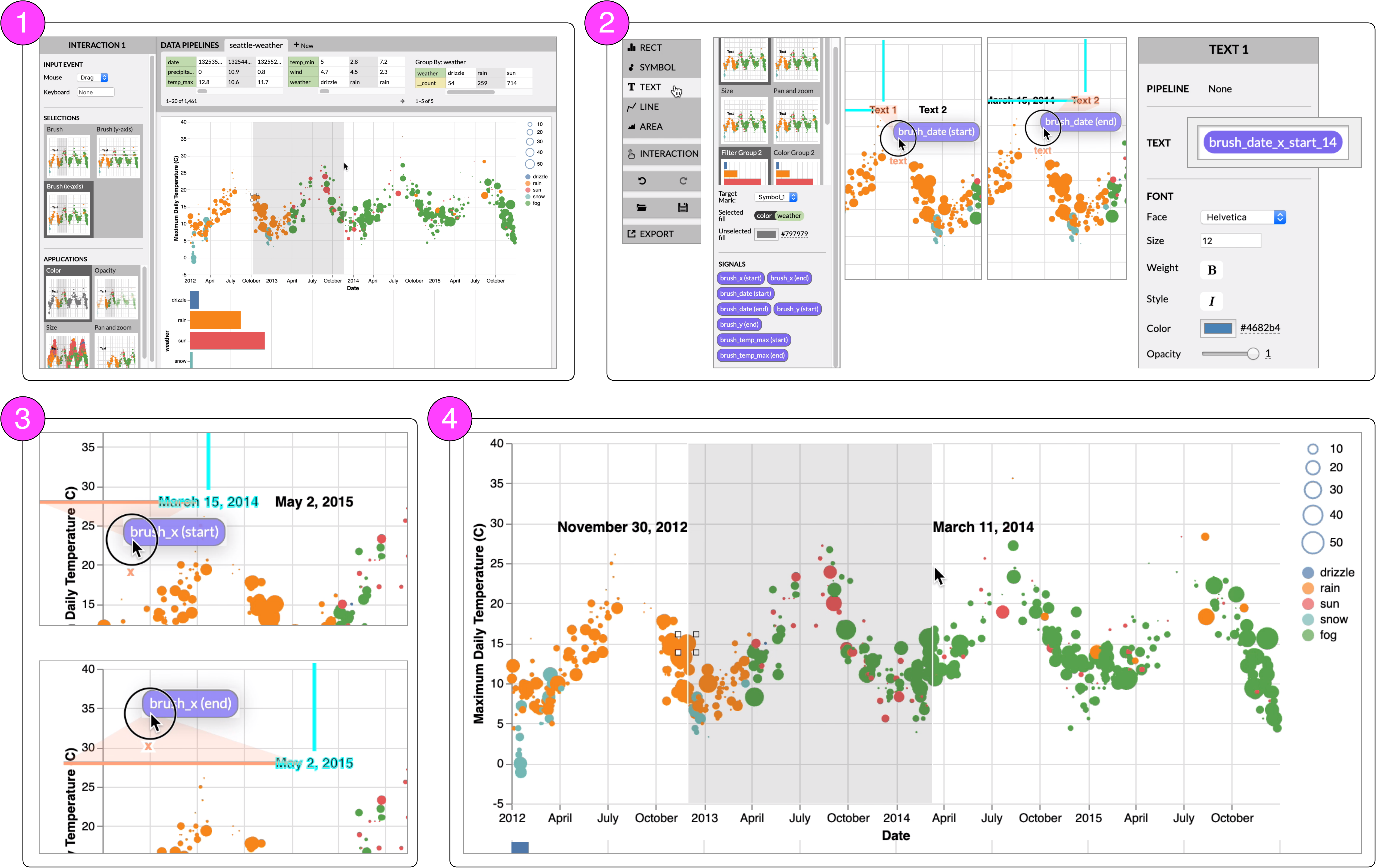}
  \vspace{-3mm}
  \caption{Brush with labeled extents. (1) A brushing interaction authored via demonstrations (see Fig.~\ref{fig:walkthrough}) (2) Binding signals that represent the brush's start and end extents to the content of two text marks respectively. (3) Positioning the text marks by their horizontal position to the brush's start and end x-coordinates. (4) The completed design: an interval selection with labeled extents.}
  \label{fig:extents}
  \vspace{-5mm}
\end{figure*}

\section{Interaction Design by Demonstration in Lyra 2}

Lyra 2's interaction design by demonstration approach comprises two parts: an abstract model for representing interaction techniques, and its articulation in graphical user interface components.
In this section, we first walk through a non-trivial usage scenario that illustrates a user\replace{ journey through}{'s process with} the interface, and then explain \replace{the}{Lyra 2's} system design\del{ that enables interaction design in }.

\subsection{Usage Scenario Walkthrough}

To illustrate the expressiveness and usability of interaction design in Lyra 2, we walk through the process of recreating \emph{Seattle Weather Exploration}~\cite{vega-seattle-weather}, an example interactive visualization from the Vega-Lite example gallery (Figure~\ref{fig:teaser}).
In this multi-view visualization, users can brush in the scatterplot to re-aggregate the histogram, and click histogram bars to highlight corresponding points in the scatterplot.

Users start by clicking the \emph{Add Interaction} button on the toolbar (Fig.~\ref{fig:walkthrough}(a)(1)), which adds an interaction specification to Lyra's state and opens the corresponding property inspector in the left-hand sidebar.
While an interaction inspector is open, the system treats user inputs on the canvas as interaction demonstrations.
As the user drags on the scatterplot (Fig.~\ref{fig:walkthrough}(a)(2)), the system uses heuristics to populate the inspector with suggestions for possible interpretations of that demonstration. 
These suggestions are grouped into two categories: \emph{selections} and \emph{applications}. 
Selections determine how input events map to a set of data tuples, while applications describe how selections drive the properties of visual elements (e.g., conditional encoding or filtering).
For each\del{ selection or application} suggestion, a thumbnail previews how the visualization would behave if the suggestion were applied (Fig.~\ref{fig:walkthrough}(a)(3)).
\replace{The user clicks the preview for a rectangular brush in the x-axis to constrain their selection, and}{If the user drags with a horizontal trajectory, the system infers a rectangular brush selection constrained in the x-axis. In the inspector, the user} clicks the \emph{color} and \emph{filter} applications to highlight selected points while filtering the data of the histogram. This enables the desired brush interaction, which is immediately active on the visualization canvas (Fig.~\ref{fig:walkthrough}(a)(4)).

To enable interactivity on the histogram, the user initializes another interaction using the toolbar (Fig.~\ref{fig:walkthrough}(b)(1)), and demonstrates a click on a histogram bar (Fig.~\ref{fig:walkthrough}(b)(2)). 
The system populates the sidebar with selection and application suggestions but, unlike the case of dragging, suggests selections on \emph{points} rather than \emph{intervals}.
The user intends to filter the scatterplot for points matching the selected bar's \texttt{weather} field. 
To match on this field, the user chooses a \emph{projected} selection in the inspector. The system automatically infers the \texttt{weather} field because it is used in a visual encoding, therefore projecting on it is likely to be meaningful.
The user once again chooses the color and filter applications to highlight the selected bar while filtering the data of the scatterplot (Fig.~\ref{fig:walkthrough}(b)(3)). Updates to the interaction specification are immediately reflected in the visualization canvas (Fig.~\ref{fig:walkthrough}(b)(4)).

At this point, the user has recreated the Vega-Lite example using only a few clicks and drags, and no textual specification. What if the user wants to more precisely label their brush extent to exactly specify the selected range? Because the brush selects a \texttt{date} range, its coordinates are continuous and may not map precisely to any tuples in the dataset. Vega-Lite cannot express the brush extent labels without this direct mapping, but the Lyra inspector exposes a set of lower-level Vega signals that can be dragged and dropped onto mark properties. 

After the user demonstrates a brush interaction (Fig.~\ref{fig:extents}.1), the property inspector will surface a list of signals related to the brush for both geometric and data coordinates. The user creates two text marks representing the start and end extent labels. They drag the \texttt{brush\_date (start)} signal onto the text content dropzone to bind the value of the signal to the mark's content, and do the same for the \texttt{brush\_date (end)} signal (Fig.~\ref{fig:extents}.2). Using a similar drag and drop process, the user binds the \texttt{brush\_x} start and end signals to the mark's x-position (Fig.~\ref{fig:extents}.3). These signals update the position and content of the labels as the user performs brush interactions on the visualization (Fig.~\ref{fig:extents}.4).

\add{When the user is ready to share their work, they click the \textit{Export} button to download the interactive visualization as a Vega specification.}


\subsection{The Abstract Model for Interaction Designs}
\label{sec:model}

\begin{figure*}[t]
  \centering
  \includegraphics[width=0.95\linewidth]{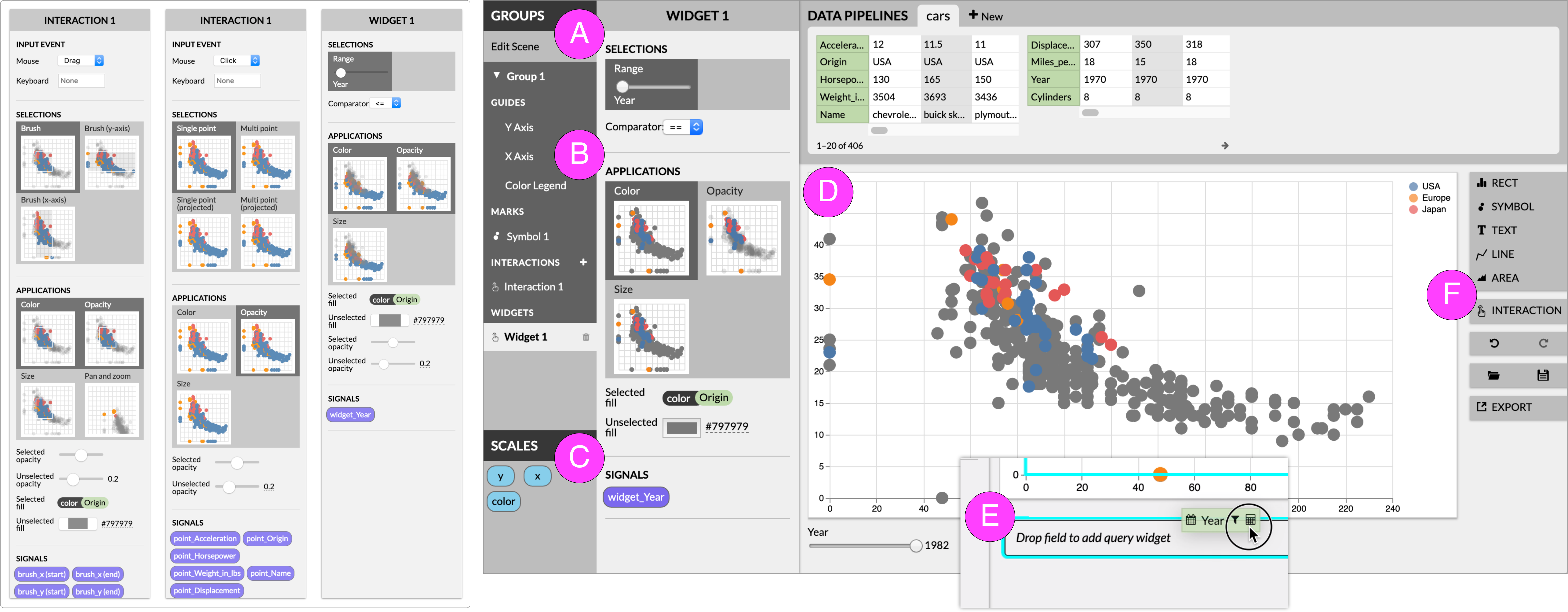}
  \vspace{-3mm}
  \caption{Lyra 2's interaction by demonstration interface. The interaction property inspectors (left) display suggestions for (a) \emph{Selections} and (b) \emph{Applications}. They also expose (c)  Vega signals to enable custom interactions. (d) Users can directly demonstrate interactions onto the visualization canvas, which reflects the current state of the output visualization. To create query widgets, users drag and drop fields onto (e) the widget dropzone. The (f) \emph{Add Interaction} button initializes interaction definitions, akin to the nearby \emph{Add Mark} buttons.}
  \label{fig:demonstration_interface}
  \vspace{-5mm}
\end{figure*}

In Lyra 2, users specify interaction techniques as a set of \emph{selections} and \emph{applications}, and can also construct dynamic query widgets~\cite{ahlberg1992dynamic}. 
This interaction model draws on abstractions found in Vega and Vega-Lite but makes some key departures. 
These differences reflect the different affordances of textual versus graphical user interfaces: Vega and Vega-Lite primitives are designed to compose together easily, to minimize language surface area and complexity; Lyra 2, on the other hand, is more concerned with recognition over recall~\cite{lidwell2010universal} by providing consistent user introspection into primitives via property inspectors.

\subsubsection{Selections}

Lyra 2's selections determine how input events select data records. They can be one of two types: points or intervals. Unlike Vega-Lite, our selection model groups selections by input event\,---\,point selections for clicks and other discrete events, and interval selections for drags. This diverges from Vega-Lite's single, multi, and interval selections because its purpose is to enable demonstration from user input events.

\replace{Like}{As with} Vega-Lite\del{ selections}, the inclusion criteria for Lyra's 2 selections can be modified via projections.
However, Lyra 2 once again diverges: \replace{projections for interval selections}{interval selection projections} (i.e., single dimensional brushes along the x- or y-axis) can be specified via demonstration and are automatically surfaced as suggestions; point selection projections (i.e. selecting a point and all other points that match its value in a given field) use the demonstration to suggest a default field to project, which can be changed via property inspector.
This difference \del{between interval and points }is due to formative feedback\del{ we received} from users during our design process: while it was straightforward to demonstrate a single-dimensional interval by moving the mouse in roughly only the horizontal or vertical direction, a similar interaction for point selection was too ambiguous.
Users would have to repeatedly click several points in order for the demonstration heuristics to infer shared field values, resulting in a frustrating experience. 

Critically, some interactive behaviors cannot be defined in terms of sets of selected records. 
For instance, consider labeling the corners of a brush: these are arbitrary coordinates in data space, and may not\del{ precisely} map to specific data records.
Such an interaction design is not expressible in Vega-Lite but, in Lyra 2, users can unwrap selections into their constituent Vega signals (Figure~\ref{fig:extents}): interval selections expose signals for the selection's start and end extents, and point selections offer signals for the selected point's backing data values. Point selections that respond on hover also include signals for mouse position.
These signals can then be dragged and dropped, akin to data fields, \del{onto dropzones or property inspectors} to establish conditional encodings or drive data transformation operators.


\subsubsection{Applications}

In Lyra 2, the application of selections to other constructs (e.g., driving conditional mark encodings, scale domains, or data transformation) is treated as a first-class primitive.
Applications reference a source selection and a target element, and are a salient departure from the Vega-Lite interaction model.
In Vega-Lite, applying selections to marks, scales, and data transformations involve subtly different syntax. 
For example, panning and zooming is implemented via a \emph{bind} transformation specified as part of a selection's definition while conditional encoding logic is inline, as part of a mark's specification.
Such distinctions would seem arbitrary and confusing within a graphical interface.
Instead, in Lyra 2, selection applications abstract over these distinctions and are surfaced as sibling suggestions during a demonstration.
Critically, by treating applications as first-class primitives, we are able to surface applications from two points of view: in a selection's property inspector, we can list all the ways it is applied across the visualization; and from the individual target elements, property inspectors update to reflect their interactive nature. 
By contrast, the former affordance is not available in Vega-Lite. Users are forced to search through a specification in order to understand the various effects a selection may have on the visualization.


\subsubsection{Query Widgets}

Both Vega and Vega-Lite support query widgets through similar mechanisms: signals and selections, respectively, can be bound to HTML input widgets like textboxes, radio buttons, and range sliders. 
Lyra 2 treats query widgets distinctly from selections for two reasons.
First, as described in the next subsection, suggestions for query widgets require different heuristics\,---\,namely, using the measure type of the bound data field (e.g., nominal, quantitative, etc.) rather than the event type of a user's demonstration.
Second, it allows us to more fluidly bridge the two levels of abstraction. Query widgets can be treated as signals to directly set the property of a mark or scale, or update data transformations. But they can also act as a selection, with a customizable inclusion criterion\,---\,functionality that is not yet possible in Vega-Lite where query widgets are treated simply as an alternate way to populate a selection.
Via a property inspector, users can specify alternate (in)equality operators to determine which records the query widget selects.

\subsection{The Demonstrations Interface}

Users introspect and manipulate the abstract interaction model through new extensions to Lyra 2's graphical interface. 

\subsubsection{Visualization Canvas}

We augment the canvas to allow interaction demonstrations directly on the output visualization (Figure~\ref{fig:demonstration_interface}(d)).
For static visualizations, the Lyra canvas always reflects the current state of the output. 
To keep the \emph{gulf of evaluation}~\cite{hutchins1985direct} narrow, we sought to maintain this property when users are designing interactions.
In contrast to widely-used prototyping tools like Figma and InVision --- where users define interactions in an editor but can only test them in a separate preview mode --- the Lyra 2 canvas continues to directly reflect the current output state, including interactions.
After creating an interaction, the user can immediately interact with it in the same view, without a separate preview.
As users create more complex visualizations with multiple interactions, they can quickly understand how different interactions behave in combination.

This immediacy creates advantages for rapidly prototyping and evaluating interactions, but causes potential ambiguity in user inputs. User inputs can have three meanings: interacting with Lyra's user interface elements (e.g., buttons, drop-down menus, etc.), interacting with the output visualization (e.g., tooltips), and performing a demonstration.

During feasibility tests, we found that interacting with the output visualization and interacting with Lyra interface elements do not conflict because their effects operate in distinct spaces: interacting with the visualization only affects the visualization state, while interface interactions only affect the Lyra state. 
As a result, these effects can coexist. 
For instance, when a user clicks on a mark, the click can both trigger any point selections that have been instantiated as well as open the mark's property inspector in Lyra's interface without any issue.

Demonstrations, however, bridge between the states of the visualization and Lyra and thus can potentially conflict.
For example, when a user clicks, how should Lyra understand whether they intend to demonstrate a point selection, populate the selection, or open the mark's property inspector?
To disambiguate this type of interaction, we introduce an \emph{implicit} demonstration mode: Lyra 2 treats user input as demonstrations when an interaction's property inspector is open.
We call this mode implicit because switching into and out of it occurs as part of a user's regular use of the interface: they have either clicked the \emph{Add Interaction} button on the right-hand side toolbar (Figure~\ref{fig:demonstration_interface}(f)) or they have manually opened the property inspector using the left-hand side listing, two operations that mirror how a user would add a mark to the canvas, or edit its properties.

Consider the common scenario where a user only wants to make a static visualization. Say they click on a mark intending to select it in the inspector. If demonstration mode is always on, this event would also be interpreted as a point selection demonstration. Since there is no interaction selected in the inspector, the system would then need to create an interaction and select it in order to define the point selection. When demonstrations are responsible for both record creation and modification, inputs may contradict user intent and result in the creation of extraneous interaction specifications. 
Separating interaction creation from modification makes the mode of user input unambigious.



\subsubsection{Query Widget Dropzone}

To support constructing query widgets, we reuse Lyra's existing \emph{dropzone} metaphor: shaded regions overlaying the canvas onto which data fields can be dropped to establish a data binding.
A new \emph{widget dropzone} appears below the canvas (Figure~\ref{fig:demonstration_interface}(e)), and works in a similar fashion\,---\,to create widgets, users drag a field from the data table, and drop it over this dropzone. 
As query widgets operate over data space, this dropzone only appears once a data binding operation has occurred.



\subsubsection{Suggestions Heuristics} 

\replace{
Lyra 2 processes demonstrations and widget drops to display potential selections and applications in the inspector sidebar, where users can quickly preview suggestions and apply them to the visualization.
}
{
When the user initiates a demonstration or widget drop, Lyra 2 evaluates a system of heuristics in four phases\,---\,enumerating selection types, enumerating application types, enumerating signals, and inferring defaults based on the demonstration.
}
Heuristics take the following inputs:
the event type (click or drag), the marks and scales present in
the current view, and marks in other views that share the same data source. \add{Heuristics are currently implemented as if-then-else rules over these properties, akin to Lyra 1's scale inference production rules~\cite{satyanarayan2014lyra}. Here, we describe the intuition behind our heuristic designs, and provide a formal treatment of their implementation in supplementary material.}

\replace{From the event type,}{In the first phase,} the system \add{uses the input event type to} distinguish\del{es} between selection types (Fig.~\ref{fig:demonstration_interface}(a)): clicks produce\del{ suggestions for} point selections, while drags yield interval selections.
Once the selection type is determined, additional heuristics \del{fire to }suggest ways of customizing the selection that are meaningful based on the types of the data fields, marks, and scales participating in the current view.
\add{%
For example, if the user chooses to project a point selection, heuristics suggest \del{sensible defaults based on }the fields participating in visual encodings\,---\,for instance, as the histogram in Fig.~\ref{fig:walkthrough}(b) binds the \texttt{weather} field to a color encoding, heuristics will by default project over \texttt{weather}.%
}
\replace{For example}{Similarly}, heuristics to customize interval selections look to the spatial relationships defined within the chart. 
If the user drags on a view containing a rect, symbol, or text mark with continuous x- and y-scales, the system suggests a regular two-dimensional brush as well as brushes constrained to the x- or y-dimension.
In contrast, the same demonstration on a view containing only an area mark, or a rect mark with a discrete x-scale and a continuous y-scale over an aggregate field (i.e., a vertical histogram) will only suggest brushing along the x-axis.
\add{%
These heuristics prevent semantically incorrect interaction designs that are expressible in Vega or Vega-Lite.
For example, with either tool, users can author specifications
for brushing along the aggregate measure of a histogram\,---\,an interaction that does not capture a valid set of backing data tuples.
Lyra 2's heuristics would not suggest these types of selections, and thus users will never enter this undesired state.%
}

When a user drags a field into the widget dropzone, the system uses heuristics to suggest widgets analogously to selections. 
However, instead of an input event type, the widget heuristics use the measure type of the widget's bound field (e.g., nominal, quantitative, etc.).
For instance, the system suggests radio buttons and dropdown menus for fields with discrete data values, and sliders for continuous values.



\replace{The applications that are suggested}{In phase two, applications are enumerated} (Fig.~\ref{fig:demonstration_interface}(b)) \del{are }based on which mark types and visual encodings are currently in use.
For instance, for discrete mark types (i.e., marks besides areas or lines), the system suggests conditional color and opacity encodings; and, for symbol marks, an additional suggestion of conditional size encodings is also surfaced.
Similarly, if continuous scales are present, a suggestion for panning \& zooming is made.
And, if marks in other views share the same data source as the mark in the demonstration view, the system suggests multiview linking and crossfilter applications.

\replace{Finally}{In the third phase}, the system uses the input event type, scale definitions, and dataset fields to suggest signals for custom interactions (Fig.~\ref{fig:demonstration_interface}(c)). Interval selections will surface signals corresponding to the brush extents in x/y coordinates, and in data coordinates based on the fields referenced in the x- and y-scales. Point selections surface signals for each field of the selected data point, enabling users to create tooltips and labels. Point selections using hover events will additionally expose signals for mouse position in x/y and data coordinates. These signals are not suggested for click-based point selections, where the selected data does not depend on current mouse position.

\add{The fourth and final phase uses the demonstration to determine default choices from the available suggestions generated in phases one and two. These heuristics work by considering the demonstration's input event history. If the user has demonstrated a drag, for instance, the heuristics take the collection of events along the drag path and calculate the angle of the drag trajectory. Drags within a $30^{\circ}$ angle from the vertical will default to brushing along the y-axis (and similarly defaulting to horizontal brushes for drags within a $30^{\circ}$ angle from the x-axis). Drags that do not tend toward either axis will assign unconstrained brush selections. For point selections, we use a threshold of 800ms to chunk a series of clicks as a distinct demonstration. We initially used the threshold for double-clicks (500ms). But, after iterative prototyping, we increased the threshold to account for sparse visualizations.
If more than one click occurs within this period, Lyra 2 defaults to a multi selection, and otherwise defaults to a single selection.}

\add{Our heuristics trade off between suggestion specificity and user agency.
In particular, we prioritize continuous user insight into the system state in accordance with direct manipulation principles~\cite{hutchins1985direct} and avoid committing users to automatic suggestions without their active input~\cite{blackwell:cogdim}.
For instance, we considered more complex inferences such as voronoi tessellations to accelerate selection when users click nearby points (akin to Vega-Lite's ``nearest'' property).
However, during feasibility tests, we found the ambiguity of such demonstrations meant that it would be easy to apply this suggestion unintentionally, and users would become frustrated at having to undo an action they did not initiate. 
Instead, our heuristics are intentionally conservative and rely on visualization properties the user has explicitly defined.}

\subsubsection{Interactions Inspector}

With straightforward extensions to Lyra's interface, we list interactions (both selections and query widgets) in the left-hand sidebar. 
The interactions property inspector (Figure~\ref{fig:demonstration_interface}(left)) enables users to visually preview selections and applications generated by their demonstration, allows fine-grained control over interaction properties, and exposes lower-level signals for custom interactions.

Each suggested selection and application is shown as a preview thumbnail that visually depicts the suggestion (Figure~\ref{fig:demonstration_interface}(a,b)).
Thumbnails narrow the \emph{gulf of evaluation}~\cite{hutchins1985direct} by providing a close mapping to a user's mental model of their desired outcome, both in terms of the selection they wish to make (e.g., previewing 1D or 2D brushes) and the effect it should have on the visualization (e.g., highlighting or filtering points).
In our design process, we considered alternatives with natural language descriptions of the corresponding Vega-Lite specifications, but found that novice users were not always familiar with terms we use for interactions (e.g.\,``brushing'') despite having a clear image in their mind. 
And, even among people who have experience creating interactive visualizations, preview thumbnails abstract over differing vocabularies that tools may expose.


The inspector also exposes relevant low-level signals that can drive custom interactions (Figure~\ref{fig:demonstration_interface}(c)). For example, a drag demonstration will surface signals for the brush extents as both visual and data coordinates, while a hover interaction will expose mouse position in both spaces. We reuse Lyra's existing rounded-rect motif, to indicate that signals can be dragged and dropped across the interface, akin to data fields. Users can, for example, drop the data coordinate signal onto a text mark's content dropzone to display its current value in the visualization (Figure~\ref{fig:extents}), or drop the x-position signals onto the mark's x-dropzone to have the mark follow the mouse (Figure~\ref{fig:extents}.3).

\del{As the property inspector lists all applications of an interaction across the visualization --- including its participation in conditional mark encodings, scale domains, or data transformations --- interaction property inspectors provide a single interface element through which users can understand the effect of an interaction.
This approach is in contrast to Vega and Vega-Lite, where only the definition of an interaction is localized, and its applications are scattered through the remainder of the specification.}


  
  
  

%% file: sections/eval_examples.tex
\begin{figure*}
  \centering
  \includegraphics[width=\textwidth,height=\textheight,keepaspectratio]{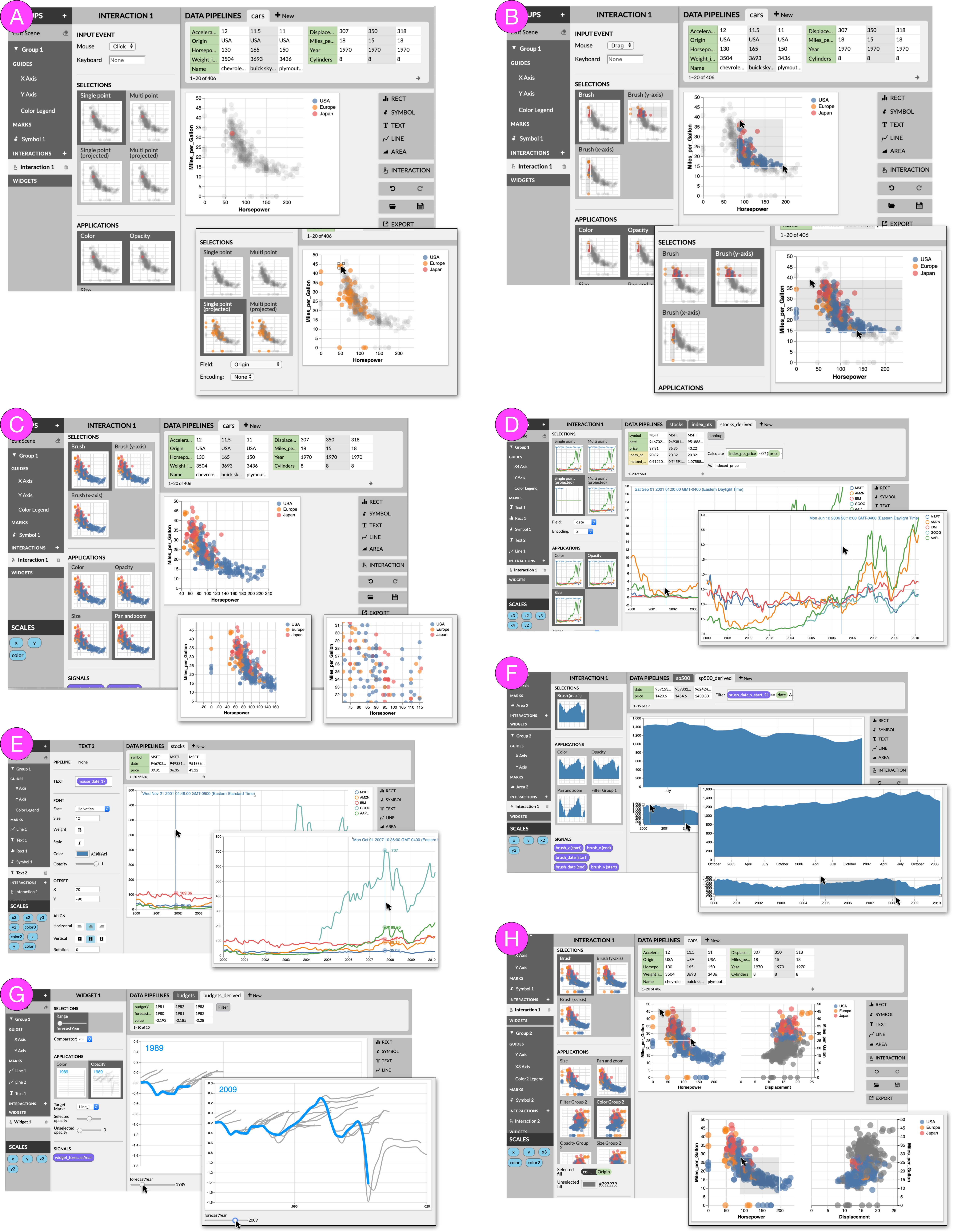}
  \caption{Example interactive visualizations demonstrating Lyra 2's coverage over Yi et al.'s taxonomy~\cite{yi:understanding}. (a, b) \emph{Selecting} marks of interest; (c) \emph{Exploring} subsets of data via pan \& zoom; (d) \emph{Reconfiguring} data via an index chart; \emph{Abstract/Elaborate} data via (e) tooltips or (f) an overview+detail visualization; (f) \emph{Filtering} data via query widgets, recreating a New York Times visualization~\cite{cox2010budget}; (g) \emph{Connecting} related tuples via brushing \& linking. \add{Walkthroughs are provided in supplementary material.}}
  \label{fig:example_gallery}
\end{figure*}

\section{Evaluation: Example Gallery}

To evaluate the expressive extent of our approach, we use Lyra 2 to create a gallery of diverse examples~\cite{ren2018reflecting}. 
These examples are drawn from the Vega and Vega-Lite example galleries and, following the corresponding papers~\cite{satyanarayan2014declarative, satyanarayan2015reactive, satyanarayan2016vegalite}, demonstrate coverage over Yi et al.'s taxonomy of interaction techniques~\cite{yi:understanding}.
In particular, as shown in Figure~\ref{fig:example_gallery}, we cover six out of the taxonomy's seven categories: we can \emph{select} marks of interest as individual points (Fig.~\ref{fig:example_gallery}(a)) or as brushes (Fig.~\ref{fig:example_gallery}(b)) with customizations inset; we can \emph{explore} different subsets via panning \& zooming (Fig.~\ref{fig:example_gallery}(c)); we can \emph{reconfigure} data, as in the case of an index chart that normalizes data based on the mouse position (Fig.~\ref{fig:example_gallery}(d)); we can \emph{abstract/elaborate} data through tooltips (Fig.~\ref{fig:example_gallery}(e)) or via an overview+detail visualization (Fig.~\ref{fig:example_gallery}(f)); we can \emph{filter} data either through direct manipulation on the visualization {(Fig.~\ref{fig:teaser})}, or via query widgets (Fig.~\ref{fig:example_gallery}(g), which recreates Amanda Cox's iconic ``porcupine chart'' from the New York Times~\cite{cox2010budget}); and, finally, we can \emph{connect} related tuples together via brushing \& linking (Fig.~\ref{fig:example_gallery}(h)).

Due to its abstract model for interaction designs (\S\ref{sec:model}), Lyra 2's expressive gamut lies between Vega and Vega-Lite. 
By being selection-based, all interaction techniques that can be constructed in Vega-Lite are expressible in Lyra 2 as well\footnote{\replace{At the time of submission, a}{A}ny differences in specifying interaction techniques between Lyra\add{ 2} and Vega-Lite are due to limitations in Lyra\add{ 2}'s support for static visualization design including the lack of a binning\del{ data} transform \replace{and no}{or} support for cartographic projections.}. 
\replace{And, b}{B}y treating query widgets as distinct from selections, and by exposing appropriate signals for each selection, Lyra begins to move beyond Vega-Lite in two key ways.
First, this expands Lyra 2's expressive extent, enabling interactive designs that are not possible in Vega-Lite including using inequality comparators for query widgets (Fig.~\ref{fig:example_gallery}(g)) or directly encoding signals values (Fig.~\ref{fig:extents}).
Second, some interactive designs are constructed more performantly in Lyra 2. 
Consider the vertical rules in the index chart and tooltip examples (Figs.~\ref{fig:example_gallery}(d, e)).
In Vega-Lite, the only way to dynamically position this rule is by applying a selection to filter the backing dataset; unfortunately, this also produces one rule per symbol for a total of five rules overlaid on top of each other.
In Lyra 2, we only need one rule and bind its \texttt{x} position to the selection's signal directly.

\emph{Limitations}. Lyra 2 does not yet support designing interactivity that is not selection-based.
Such techniques primarily fall within Yi et al.'s \emph{Encode} category, which describes interactive behaviors that bypass data space and manipulate the view directly (e.g., changing the mark type, which visual channels are encoded, or which data fields participate in visual encoding).
Similarly, although Lyra 2 supports HTML widgets, it does so as \emph{query} widgets\,---\,i.e., these widgets manipulate expressions in data space and cannot be used to modify properties of marks directly.
Finally, Lyra 2 does not expose the full expressive power of Vega's signals; instead, only signals that pertain to selections are available to users.
As a result, more complex and custom selection-based techniques (e.g., reordering the dimensions of a matrix or DimpVis~\cite{kondo2014dimpvis}) remain out of Lyra 2's range.
How to enable more expressive selection-based and non-selection based interactive techniques in a graphical and direct manipulation medium is a compelling direction for future work.

  
  
  
  
  

%% file: sections/eval_study.tex
\section{Evaluation: First-Use Study}

We designed Lyra 2 to improve expressiveness and usability for users, especially those with less prior coding experience. We evaluate our approach's usability through a first-use study with 6 representative users including 2 experienced visualization designers, 2 computer scientists, and 2 from fields unrelated to visualization. The mean self-reported past visualization design expertise was 2.83 on a 5 point Likert scale ($\sigma = 1.21$). Participants with prior experience had primarily used D3, with only one participant familiar with the Vega stack.

\subsection{Methods}

We began each study with a 10 minute walkthrough of Lyra 2's features. We then asked participants to complete three interaction design tasks. For each task, we showed them an example interactive visualization and created the static version in Lyra 2. We then asked them to use Lyra 2 to recreate the interactivity from the example. The three visualizations were drawn from standard Vega-Lite examples: a pan and zoom scatterplot (\replace{T1}{$T_{panzoom}$}), a filterable scatterplot with query widgets (\replace{T2}{$T_{widgets}$}), and a linked scatterplot and bar chart (\replace{T3}{$T_{linked}$}). These tasks were designed to maximize participant engagement with Lyra 2's interaction design interfaces, and were ordered in increasing difficulty. Participants were encouraged to think aloud as they completed the tasks. At the end of the study, we asked each participant to rate the usefulness of each part of the Lyra 2 interface on a 5-point Likert scale. We also set aside open-ended conversation time for participants to explore the tool, ask us questions, and share reflections on exciting or challenging aspects of completing the tasks. Sessions lasted approximately 45 minutes, and participants were compensated with a \$15 \add{Amazon} gift card\del{ to an online retailer}.

\subsection{Quantitative Results}

Users quickly learned how to create interaction designs in Lyra 2, and all users, regardless of their prior visualization experience, successfully completed all three tasks with minimal guidance. The average task completion times for the tasks were, in minutes and seconds, \replace{T1}{$T_{panzoom}$}: ($\mu = 1\colon07, \sigma = 1\colon07$), \replace{T2}{$T_{widgets}$}: ($\mu = 1\colon32, \sigma = 0\colon39$), \replace{T3}{$T_{linked}$}: ($\mu = 8\colon51, \sigma = 2\colon19$). In a post-study survey, participants rated Lyra 2’s demonstrations interface highly: demonstrating interactions on the canvas felt natural ($\mu = 4.83, \sigma = 0.37$ on a 5-point Likert scale), suggestions were useful ($\mu = 4.83, \sigma = 0.37$), and previews helped evaluate suggestions ($\mu = 4.17, \sigma = 1.07$). 

\subsection{Qualitative Results}

Participants found the interaction design by demonstration process\,\emph{\replace{"}{``}natural\replace{"}{''}}. An experienced participant said that the user flow was similar enough to their development process in textual specification tools that they could easily transfer their skills. Less experienced participants also found demonstration's short articulatory distance helpful. Thinking aloud during \replace{T3}{$T_{linked}$}, one participant said,\,\emph{``Demonstrating the interactions was very easy. I didn't really know the word brushing, but it was easier to just do it than to say what it is. Same with picking from the widget types that have technical names like radio.''}

Participants were especially excited about easily creating interactions they considered complex. A participant said that Lyra 2\,\emph{``would make me experiment more with possible interactions''}\,because of the lower technical barrier. Reflecting on \replace{T3}{$T_{linked}$}'s multi-view filtering, they said,\,\emph{``I wouldn't have thought to make this. I would think it was too hard.''}\,Another less experienced participant noted that easily creating multi-view filtering would be very useful in their work, where people often use non-interactive charts of high dimensional data.

The quick feedback loop of evaluating and applying suggestions also stood out positively to participants. A more experienced participant compared the feedback loop with textual specification, saying that\,\emph{``with these lower level libraries, getting an interaction to work takes a while even when copying and pasting.''} Similarly, a participant with less visualization experience noted that\,\emph{``being able to test immediately was very useful. Even when it went wrong, I could immediately tell that it was wrong.''}\,Previews were important to this quick feedback loop, but users noted that not all of them were equally useful. For instance, by default, previews for single and multi point selections look the same which led users to be unsure about how multi point selections differ.


The primary shortcoming we observed was when participants' mental models did not match Lyra 2's interface.
For example, one participant pointed out that, although panning and zooming is a drag interaction,
the interface forced them to first choose a selection using the ``brushing'' terminology; they might have instead expected to select panning \& zooming directly.
Similarly, a few participants were initially unsure whether they should demonstrate on the source or target view to surface a multiview filter suggestion. 

\replace{While reviewing study recordings, o}{O}ne participant's question during the post-study debrief struck us as particularly insightful:\,\emph{``what if I want to make interactions that aren't in the suggestions?''}\,This question suggests a drawback to our approach\del{ that} we had not previously considered: might suggestions limit what users consider to be expressible in Lyra 2?
\replace{Indeed, t}{T}his concern does not appear particular to Lyra 2 or \del{the domain of }interaction design by demonstration.
For example, \add{in }prior studies of mixed-initiative systems\replace{ have reported}{,} users of Voyager worr\replace{ying}{ied} about whether its visualization recommendations might cause them to\,\emph{``start thinking less''}~\cite{2017-voyager2} and users of an interactive machine translation system perceiv\replace{ing}{ed} themselves\,\emph{``less susceptible to be creative''}~\cite{2014-ptm}.
\replace{Thus, o}{O}ur results add further evidence that better balancing agency and automation~\cite{heer2019agency} is a critical avenue for future work.

%% file: sections/eval_cognitive.tex
\section{Evaluation: Cognitive Dimensions of Notation}
\label{sec:cogdim}

In this section, we compare Lyra 2's usability to textual specification of interaction designs in Vega or Vega-Lite.
To do so, we adopt the Cognitive Dimensions of Notation~\cite{blackwell:cogdim}, a
heuristic evaluation framework that has previously been used to evaluate HCI toolkits~\cite{ledo2018evaluation} as well as visualization systems~\cite{satyanarayan2014lyra, satyanarayan2014declarative}.
Of the 14 dimensions found in the framework, we find particularly salient differences for the following:

\emph{Closeness of Mapping.} Demonstrations offer a much closer mapping between the notation of the specification (input events) and the desired outcome (an interaction design). Depicting suggestions as thumbnails further builds on this dimension, by offering users a visual preview of possible interactive behaviors. By contrast, textual specification languages force users to express interaction techniques in potentially unfamiliar terms. In fact, in formative evaluations, many novice users had never previously described interactions as ``selections'' or ``brushes,'' which are common terms in data visualization and HCI literature.

\emph{Progressive Evaluation} and \emph{Premature Commitment}. It is difficult to validate in-progress work with textual languages as only complete specifications produce working output. If required properties are left underspecified, for instance, the language compiler will throw an error and produce no output.
This issue is exacerbated for interaction specification: complete definitions of signals or selections will produce working output, but this may not always be evident until they are used in the remainder of the specification.
By contrast, Lyra 2's demonstrations exemplify support for these dimensions: users are able to explore the possible design and easily preview individual design choices \emph{before} explicitly instantiating a full interaction technique.
However, there is still room for further improvement: Lyra 2 is currently only able to make multiview suggestions if secondary views have already been created; recommending multiview visualizations is an active area of research~\cite{moritz2018formalizing, qu2017keeping} and future versions of Lyra should consider how to incorporate it in the context of interaction suggestions.

\emph{Diffuseness.} Textual languages, particularly a higher-level grammar like Vega-Lite, offer a much more concise specification format than the graphical equivalent in Lyra 2.
This property is true not only in the general sense (localized, often one-word changes in Vega-Lite translate to multiple clicks in Lyra 2) but also in specific ways for this paper's contribution.
By definition, demonstrations are a more ambiguous specification format and a user may have to perform several attempts before the system correctly infers their desired behavior. 
We observed this issue most saliently when attempting to project a point selection: users would need to repeatedly click several points for the system to have sufficient data to infer shared field values, which proved to be an overly frustrating experience.
Based on these results, we chose to expose point selection projections in the property inspector rather than via demonstrations.
Lyra 2 has an additional source of diffuseness: it is not difficult to imagine preview thumbnails becoming unwieldy as users craft more complex multi-view dashboards.
Future work must consider how to scale the suggestion previews\,---\,for example, once the visualization's dimensions cross a threshold, perhaps suggestions switch from purely visual to a combination of visual and textual modalities. 
However, such a change may trade off \emph{closeness of mapping}. 

\del{\emph{Abstraction.} Lyra previously bridged the Vega and Vega-Lite levels of abstraction for static visualization design, and this property continues to be true for interaction design as well.
Although users can choose to compile Vega-Lite specifications and edit the output lower-level Vega specification, few users do so in practice because of the complexity cliff.
For instance, a default Vega-Lite interval selection generates 14 lower-level signals, each with a non-trivial definition.
In Lyra 2, we are able to offer a smoother abstraction gradient: only signals corresponding to the x- and y- extents are exposed, and by dragging and dropping interactions within property inspectors, Lyra 2 is able to infer whether the interaction should be treated as a selection or a signal.}


\emph{Hidden Dependencies.} Property inspectors allow us to reveal dependencies that are otherwise more latent in textual specifications.
In particular, as we designed interaction property inspectors, we realized that they provided a prime location for collating all the ways an interaction technique may be used across a visualization. 
Working through this design motivated raising selection applications to be a first-class primitive in our interaction model. 
In the textual languages, a user would have to search through a specification and manually build their mental model of how a selection or signal is being used.

In summary, \del{using the Cognitive Dimensions of Notation framework, we believe} Lyra 2's demonstrations compare favorably to textual specification in terms of closeness of mapping, progressive evaluation and premature commitment\replace{, and offer a smoother abstraction gradient. Demonstrations, however,}{ but} result in a more diffuse user experience. 

%% file: sections/conclusion.tex
\section{Conclusion and Future Work}

This paper contributes methods for designing interactive visualizations by demonstration and instantiates these methods in Lyra 2. 
Its interface components, such as the visualization canvas with demonstrations, suggestion heuristics, and interaction inspector, narrow the gulfs of execution and evaluation for interaction designs. A diverse example gallery demonstrates Lyra 2’s expressiveness, including many designs that are nontrivial to express in current declarative visualization languages. Participants in a user study found that the tool helped them create visualizations that previously felt too difficult to attempt. Lyra 2 is available as open-source software at \url{https://github.com/vega/lyra}.


Lyra 2 represents only the first step in developing non-textual mechanisms for authoring interactivity in data visualizations, and there are several promising next steps to explore.
How to support designing more custom interactions by demonstration, especially those that are not selection-based (e.g., \emph{Encode}-type techniques~\cite{yi:understanding}), is a clear next step.
It is not clear that demonstrations can or should target low-level expressions (e.g., Vega signals) directly.
Rather, there appears to be the need for novel approaches which occupy a middle ground between direct demonstration and visual programming interfaces (e.g., InterState~\cite{oney2014interstate}).
Even with selection-based interactions, future work should consider how to go beyond heuristics~\cite{saket2018beyond} and utilize recommendation methods including ranked enumeration~\cite{mackinlay1986automating, moritz2018formalizing, wongsuphasawat2016towards} and learned models~\cite{hu2019vizml}.
A key challenge here is that these alternate approaches are grounded in empirically-validated principles for \emph{effective} visual encoding, and similar results do not yet exist for interaction design. 

Despite prior work on depicting the runtime behavior of interactive visualizations~\cite{hoffswell2016debugging, hoffswell2018augmenting}, we did not currently find a need to offer strategies for debugging interaction techniques in Lyra 2.
We believe the reason is because Lyra's graphical interface mediates user manipulations of the underlying Vega specification.
In particular, as discussed in \S\ref{sec:cogdim}, Lyra's interface surfaces a number of hidden dependencies latent in the corresponding textual specifications and, through its heuristics and the suggestions it surfaces, constrains the allowable state space\,---\,two issues that prior work has used visual debuggers to ameliorate~\cite{hoffswell2016debugging}.
Nevertheless, as future research explores designing more complex interaction techniques, the need for a debugger may once again arise.

Finally, and perhaps most critically, our first-use studies provide additional evidence for the need to better balance automated suggestions and user autonomy and agency in mixed-initiative interfaces~\cite{heer2019agency}.
Studying these issues in the domain of design may be particularly viable as new systems can leverage prior results from cognitive psychology into the role of examples in the design process~\cite{ward1994structured} and when they most spur creative work~\cite{kulkarni2014early}.
However, such systems must \add{also} grapple with the consequences of imposing theories of semantic (rather than purely syntactic) validity on design\,---\,what are the implications of systems encoding notions of ``good'' visualization~\cite{correll2019ethical, kennedy2016work}?